\documentclass[
 reprint,
 superscriptaddress,
 amsmath,amssymb,
 aps,
 prb,
 floatfix,
]{revtex4-2}

\usepackage{graphicx, epstopdf}
\usepackage{dcolumn}
\usepackage{bm}
\usepackage[colorlinks,linkcolor=blue,hyperindex,CJKbookmarks]{hyperref}
\makeatletter
\begin{document}
\title{Orbitally selective resonant photodoping to enhance superconductivity}

\author{Ta Tang}
\affiliation{Department of Applied Physics, Stanford University, California 94305, USA.}
\author{Yao Wang}
\affiliation{Department of Physics and Astronomy, Clemson University, Clemson, South Carolina 29631, USA}
\author{Brian Moritz}
\affiliation{Stanford Institute for Materials and Energy Sciences, SLAC National Accelerator Laboratory, 2575 Sand Hill Road, Menlo Park, California 94025, USA.}
\affiliation{Department of Physics and Astrophysics, University of North Dakota, Grand Forks, ND 58202, USA.}
\author{Thomas P. Devereaux}
\affiliation{Stanford Institute for Materials and Energy Sciences, SLAC National Accelerator Laboratory, 2575 Sand Hill Road, Menlo Park, California 94025, USA.}
\affiliation{Department of Materials Science and Engineering, Stanford University, Stanford CA 94305.}
\affiliation{Geballe Laboratory for Advanced Materials, Stanford University, Stanford, CA 94305, USA.}

\begin{abstract}
Signatures of superconductivity at elevated temperatures above $T_c$ in high temperature superconductors have been observed near 1/8 hole doping for photoexcitation with infrared or optical light polarized either in the CuO$_2$-plane or along the $c$-axis. While the use of in-plane polarization has been effective for incident energies aligned to specific phonons, $c$-axis laser excitation in a broad range between 5 $\mu$m and 400 nm was found to affect the superconducting dynamics in striped La$_{1.885}$Ba$_{0.115}$CuO$_4$, with a maximum enhancement in the $1/\omega$ dependence to the conductivity observed at 800 nm. This broad energy range, and specifically 800 nm, is not resonant with any phonon modes, yet induced electronic excitations appear to be connected to superconductivity at energy scales well above the typical gap energies in the cuprates. A critical question is what can be responsible for such an effect at 800 nm? Using time-dependent exact diagonalization, we demonstrate that the holes in the CuO$_2$ plane can be photoexcited into the charge reservoir layers at resonant wavelengths within a multi-band Hubbard model. This orbitally selective photoinduced charge transfer effectively changes the in-plane doping level, which can lead to an enhancement of $T_c$ near the 1/8 anomaly.
\end{abstract}

\maketitle

\section{Introduction}
Actively controlling states of matter is of great theoretical, experimental, and practical interest, particularly in the field of quantum materials, where the entangled degrees of freedom hinder the use of more reductive, straightforward control knobs. The development of laser techniques has enabled the precise control of material properties in a predictive manner and has led to the realization of new states without equilibrium analogs\,\cite{doi:10.1080/00018732.2016.1194044, doi:10.1146/annurev-matsci-070813-113258, doi:10.1146/annurev-conmatphys-031218-013423, Wang:2018aa}. 
For example, Floquet engineering of quantum materials via a periodic drive\cite{doi:10.1146/annurev-conmatphys-031218-013423} can be used to modify band topology\cite{PhysRevB.79.081406, PhysRevB.84.235108, Lindner:2011we, Rechtsman:2013wi, Sentef:2015we, Claassen:2016ti} and to induce transient dynamics in strongly correlated systems by modifying the underlying Hamiltonian\cite{PhysRevLett.115.075301, Mentink:2015vz, PhysRevLett.116.125301, RevModPhys.89.011004, PhysRevB.96.085104}.
Alternatively, one can use transient fields to alter the balance between competing orders or induce meta-stable phases\cite{Wang:2018aa}. Experimentally, pulses of light have been successfully applied to control quantum magnets\cite{Perfetti_2008, Liu:2012tu}, charge density waves\cite{Schmitt1649, Stojchevska177, Zhang:2016vf}, and excitonic order\cite{PhysRevLett.119.086401, PhysRevLett.119.247601}.

Among these applications of nonequilibrium techniques, light-induced, or light-enhanced, superconductivity has been one of the most exciting discoveries\cite{Mitrano:2016aa, Fausti:2011wa, Hu:2014ab, PhysRevB.89.184516}. 
It has been suggested that in the cuprates a transient improvement of superconducting properties - specifically the optical response - far above equilibrium $T_c$ has been realized experimentally either by selectively driving phonon modes\cite{PhysRevB.89.184516, Hu:2014ab} or by applying near-infrared optical pulses\cite{PhysRevB.90.100503, PhysRevB.91.174502, PhysRevX.10.011053}. However the actual microscopic cause of the effect is still under debate.

Non-equilibrium techniques can provide unprecedented access to the excited-state manifold to assist in unraveling the origins of unconventional superconductivity itself. In equilibrium, superconductivity emerges from the insulating cuprate parent compounds by varying doping and temperature, generically forming a dome-shaped region in the rich phase diagram\cite{RevModPhys.78.17}. Despite many years of study, much of the underlying physics of the cuprate equilibrium properties remains unclear\cite{Keimer:2015aa}.
Making the situation more daunting, the superconducting phase generally has a complex relationship with other phases, such as charge stripe order\cite{Keimer:2015aa, RevModPhys.87.457}. For example, in YBa$_2$Cu$_3$O$_{6+x}$, the charge order develops above $T_c$ but is suppressed when superconductivity emerges below $T_c$\cite{Chang:2012ty, Gerber949}. Conversely, the charge order below $T_c$ is strengthened upon applying magnetic fields\cite{Chang:2012ty, Gerber949} or pressure\cite{Kim1040}, which tend to suppress superconductivity. These results point to a complex competition between charge order and superconductivity.

The suppression of superconductivity with the rise of charge order is particularly evident in materials like $\textrm{La}_{1.8-x}\textrm{Eu}_{0.2}\textrm{Sr}_{x}\textrm{CuO}_{4}$(LESCO) and $\textrm{La}_{2-x}\textrm{Ba}_{x}\textrm{CuO}_{4}$(LBCO), where a local minimum of $T_c$ near $1/8$ doping is found, as superconductivity competes with a strong charge-stripe order\cite{PhysRevB.40.7391, MACHIDA1989192, doi:10.1143/JPSJ.59.1047, Tranquada:1995aa, PhysRevLett.78.338,PhysRevLett.88.167008}.  
It is possible that the balance between superconductivity and competing states can be shifted by applying transient optical fields, providing a route to enhance superconductivity with light. Putative light-enhanced superconductivity has been reported in several systems\cite{PhysRevLett.67.2581, doi:10.1063/1.107069, Hu:2014ab, PhysRevB.89.184516, Fausti:2011wa, PhysRevB.90.100503, PhysRevB.91.174502, Cremin19875, Mitrano:2016aa}, including transient superconductivity near the 1/8 anomaly in stripe-ordered LESCO and LBCO\cite{Fausti:2011wa, PhysRevB.90.100503, PhysRevB.91.174502}.

There are two distinct photon energy scales that have been used to photoexcite cuprates. The first uses terahertz pulses to selectively excite lattice vibrational modes\cite{Fausti:2011wa, Hu:2014ab, PhysRevB.89.184516}. Typically in these experiments, a transverse Josephson plasma resonance (JPR) appears at a temperature higher than equilibrium $T_c$ following excitation, suggesting a transiently enhanced superconducting state. 
However, the underlying mechanisms for this effect remain largely unclear. Proposed theories\cite{PhysRevB.93.144506, PhysRevB.94.155152, PhysRevLett.117.227001, PhysRevB.94.214504, Kennes:2017vp,Nava:2018wc, PhysRevB.98.214514} are based on notions that the pump field changes the interaction or structure, transforming the system into a broken symmetry superconducting state with a $T_c$ much higher than the equilibrium value. 
In LESCO near the 1/8 anomaly, the weakening of charge-stripe order has been observed after terahertz excitation tuned to an in-plane phonon mode\cite{PhysRevLett.112.157002}, 
and the enhancement of superconductivity\cite{Fausti:2011wa} may be due to the weakening of the competing stripe order.

A potentially distinct route to enhanced superconductivity makes use of $c$-axis polarized near-infrared light ($\sim 800$ nm)\cite{PhysRevB.90.100503, PhysRevB.91.174502, Cremin19875}. 
In LBCO near the 1/8 anomaly, a transverse JPR edge appears immediately following transient excitation at a base temperature of 30K, far above the equilibrium $T_c$; and a $1/\omega$ dependence in the conductivity has been observed. Moreover, when tuning the $c$-axis polarized pulse's central wavelength\cite{PhysRevB.91.174502}, the signals were observed most clearly at 800 nm. A weaker, or even null effect, was observed for pulse wavelengths at $2\,\mu \text{m}$ and $5\,\mu \text{m}$, while no sharp edge was found at 400 nm, pointing to 800 nm as being most favorable.
The 800 nm $c$-axis polarized light presumably excites the system via a dominant electronic mechanism, although the details behind such a mechanism remain unclear.  We note that numerical work suggests that an in-plane polarized pulse also can enhance $d$-wave superconducting pairing, while suppressing charge order in the vicinity of a phase boundary where competition between phases may be strong\cite{PhysRevLett.120.246402, 2021arXiv210103495W}.

In this paper, we demonstrate that $c$-axis polarized, near-infrared pump pulses can lead to an orbitally selective resonant transfer of charge from the $\textrm{CuO}_2$ plane into the charge reservoir layers via apical oxygens. This result is obtained by using time-dependent exact diagonalization (ED) which is limited to small system size and will not be able to elucidate long range orders in the system. However we believe the method is appropriate to study $c$-axis charge transfer induced by a time-dependent field, as will be justified later. While this small-cluster simulation cannot directly track the evolution of various competing orders, the quantification of the charge transfer indirectly reflects the light-manipulation of superconductivity near the $1/8$ anomaly (see FIG.~\ref{pic:lattice}). This observation presents a novel mechanism to explore the role of orbital- and material-dependent charge transfer for controlling the dynamics of holes.

\section{Model and Method}
 \begin{figure}[hpbt!]
     \begin{center}
         \includegraphics[width=8.5cm]{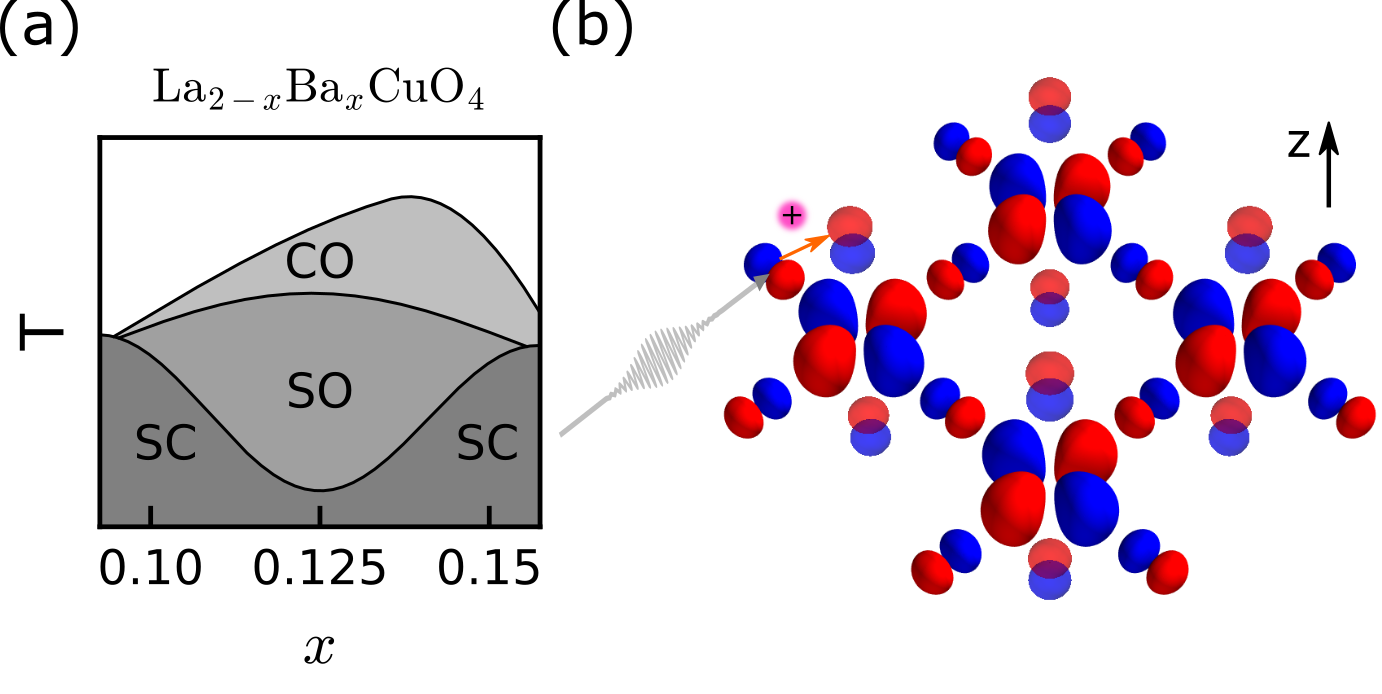}
         \caption{\label{pic:lattice}
       (a) Phase diagram near the $1/8$ doping anomaly, where there is a local minimum of $T_c$ for the superconducting phase. SO, CO and SC represent spin order, charge order and superconductivity respectively. (b) A schematic of the copper oxide plane with apical oxygen from the charge reservoir layers. The z direction is parallel to $c$-axis. The apical oxygen $p_z$ orbitals are slightly transparent.  Near-infrared photoexcitation (grey pulse) transfers charge (red dot) from the CuO$_2$ plane to the apical oxygen orbitals in the charge reservoir layers.
     }
     \end{center}
 \end{figure}

Here, we are interested in applied optical or near-infrared fields with a polarization perpendicular to the CuO$_2$ plane, where the apical oxygen atoms may play on outsized role in the photoinduced dynamics.  To model such a system, we extend a three-orbital Hubbard model of the CuO$_2$ plane by adding apical O $2p_z$ orbitals \cite{PhysRevB.43.2968} with one above ($2p_{zu}$) the copper atom and one below ($2p_{zd}$) it. The resulting multi-orbital Hubbard Hamiltonian takes the form
\begin{eqnarray}\label{eq:Hamiltonian}
\hat{H}_0 & = & \sum_{i\sigma}\left(\epsilon_d d_{i\sigma}^\dagger d_{i\sigma}\right. +\!\!\!\!\!\! \sum_{\nu\in\{x,y,z\}}\!\!\!\!\!\!\left.\epsilon_{p_\nu} p_{\nu i\sigma}^\dagger p_{\nu i\sigma} \right)+U_{d}\sum_{i}n_{d i \uparrow} n_{d i \downarrow}\nonumber\\
  & + & U_{p}\sum_{i}\!\!\!\sum_{\nu\in\{x,y,z\}}\!\!\!\! n_{p_\nu i \uparrow} n_{p_\nu i \downarrow}+U_{dp}\sum_{\left<ij\right>}\sum_{\nu\in\{x,y\}}\!\!\!\!\! n_{d i} n_{p_\nu j}\nonumber\\
  & + & \sum_{\left<ij\right>\sigma}\sum_{\nu\in\{x,y\}} \left( t_{dp_\nu}d_{i\sigma}^\dagger p_{\nu j \sigma}  + t_{p_zp_\nu} p_{z i \sigma}^\dagger p_{\nu j \sigma}+h.c.\right)\nonumber\\
  & + & \sum_{\left<ij\right>\sigma}\left(t_{p_x p_y} p_{x i \sigma}^\dagger p_{y j \sigma} + h.c.\right).
\end{eqnarray}
Here, $d_{i\sigma}$, $p_{xi\sigma}$, $p_{yi\sigma}$ and $p_{zi\sigma}$ ($d^{\dagger}_{i\sigma}$, $p^{\dagger}_{xi\sigma}$, $p^{\dagger}_{yi\sigma}$ and $p^{\dagger}_{zi\sigma}$) are annihilation (creation) operators for holes in Cu $3d_{x^2-y^2}$ and O $2p_x, 2p_y, 2p_z$ orbitals at site $i$ with spin $\sigma$, respectively; $n_{\mu i \sigma}$ is the number operator for orbital $\mu$ with spin $\sigma$ at site $i$; $\epsilon_{\mu}$ are the site energies corresponding to orbital $\mu$; $t_{\mu\nu}$ represents the hopping integral between orbitals $\mu$ and $\nu$, here restricted to the nearest-neighbor unit cells; $U_{\mu}$ denotes the onsite Coulomb interaction for orbital $\mu$ and $U_{dp}$ denotes a Coulomb interaction between the planar Cu $3d_{x^2-y^2}$ and O $2p_{x/y}$ orbitals.

To simulate the nonequilibrium dynamics and excited-state spectrum, we use the time-dependent exact diagonalization (ED) method\,\cite{RevModPhys.66.763, doi:10.1137/1.9780898719628,doi:10.1063/1.451548,doi:10.1137/S00361445024180,2005AIPC..789..269M,Saad:1992:AKS:131344.131358,doi:10.1137/S0036142995280572}. To keep the problem computationally tractable, we consider a minimal $2\times 2$ cluster (Cu$_4$O$_{16}$) with periodic boundary conditions (see FIG.~\ref{pic:lattice} (b)) and the simulation is performed at zero temperature.

Using the ground state obtained from ED as the initial state, we then perform time evolution of this initial state after modifying the Hamiltonian to account for an applied pump field. 
The electromagnetic field enters the Hamiltonian through a Peierls substitution
$
t_{ij}\rightarrow t_{ij}e^{i\int_{\mathbf{r}_i}^{\mathbf{r}_j} \mathbf{A}(t, \mathbf{r'})\cdot d\mathbf{r'}}
$.
For time evolution $\left|\Psi(t+\delta t)\right>\! =\! e^{-\frac{i}{\hbar}H(t)\delta t} \left|\Psi(t)\right>$, we employ the Krylov subspace method.
For octahedral structures (with two apical oxygens per copper atom), there is a mirror symmetry about the $x$-$y$ plane. 
As the $c$-axis pump field $A(t)$ couples to a $z$-direction ($c$-axis) current operator $J_z$ and changes the parity, it imposes a selection rule for $c$-axis optical excitations.

We first consider the equilibrium spectra for the cluster, including the single-particle spectral function $A(k,\omega)$ and the optical conductivity $\sigma_{zz}(\omega)$. 
The orbitally resolved single-particle spectral function is defined as
\begin{eqnarray}\label{eq:spectra}
A_\mu(k,\omega) & = & -\frac{1}{\pi}\textrm{Im}\left[\left<G\right|c_{\mu \bm{k}}\frac{1}{\omega - \hat{H} + E_G + i\eta}c_{\mu \bm{k}}^\dagger\left|G\right>\right.\nonumber\\
 & + & \left.\left<G\right|c_{\mu \bm{k}}^\dagger\frac{1}{\omega + \hat{H} - E_G + i\eta}c_{\mu \bm{k}}\left|G\right>\right],
\end{eqnarray}
where $\left|G\right>$ is the ground state; and the generic hole annihilation operator with momentum $\bm{k}$ can be written as $c_{\mu \bm{k}}=\frac{1}{\sqrt{N}}\sum_j c_{\mu j}e^{i\bm{k}\cdot \bm{r}_j}$, where $\mu$ is the orbital index. 
Here, for convenience we use the notation $c_{d i}=d_i$ and $c_{p\nu i}=p_{\nu i}$, omitting the spin index.

The regular part of the $c$-axis optical conductivity is defined as
\begin{equation}\label{eq:sigma_reg}
    \sigma_{zz}^{reg}(\omega)=\frac{1}{\omega N}\textrm{Im}\left\langle G\right|\hat{J_z}\frac{1}{\hat{H}-E_0-\omega-i\epsilon}\hat{J_z}\left|G\right\rangle,
\end{equation}
where $N$ is the number of sites. Only excited states with parity opposite to that of the ground state can contribute to $\sigma_{zz}^{reg}(\omega)$. 

\section{Results}
We choose the canonical parameter set relevant for LBCO \cite{PhysRevB.43.2968,PhysRevLett.105.177401,PhysRevB.42.6268,PhysRevB.38.11358,PhysRevB.38.6650}(units in eV):
\begin{eqnarray}\label{eq:para}
\epsilon_d = 0  & \quad \epsilon_{p_{x,y}}  = 2.8 & \quad \epsilon_{p_z} = 3\nonumber\\
t_{dp_{x,y}} = 1 & \quad t_{p_x p_y} = 0.5 & \quad t_{p_z p_{x,y}} = 0.3\nonumber\\
U_{d} = 8.5 & \quad U_{p} = 4 & \quad U_{dp} = 0.6\,.
\end{eqnarray}

These model parameters lead to reasonable equilibrium properties compared with experiments\footnote{For example, in our five-band model, the effective superexchange $J$ is about $135$meV obtained from the singlet-triplet gap, or $113$meV estimated from a leading term in the perturbative expansion $J = 4t_{dp}^4/(\epsilon_p + U_{dp})^2\times[1/(\epsilon_p + U_p/2)+1/U_d]$. Although there may be some ambiguity due to the multi-orbital structure, both values are close to experimental estimates $\sim 128$ meV\cite{PhysRevLett.62.2736}.}. 
At half filling (4 holes), the hole concentrations on Cu, in-plane O and apical O are 68.0\%, 31.4\% and 0.6\% respectively; at 25\% doping (5 holes), the corresponding numbers are changed to 73.1\%, 48.0\% and 3.9\%. The doped holes prefer in-plane oxygen orbitals. Specifically, the ratio between doped hole number on in-plane oxygen and on copper is about 3.25:1, which is comparable to the result ($\sim$3:1) for a three-band model with only in-plane orbitals\cite{PhysRevB.93.155166}. These results also are close to the hole distribution estimated from x-ray absorption\cite{PhysRevLett.68.2543} and qualitatively match NMR\cite{PhysRevB.90.140504} measurements. 

\begin{figure}[t!]
  \begin{center}
    \includegraphics[width=8.5cm]{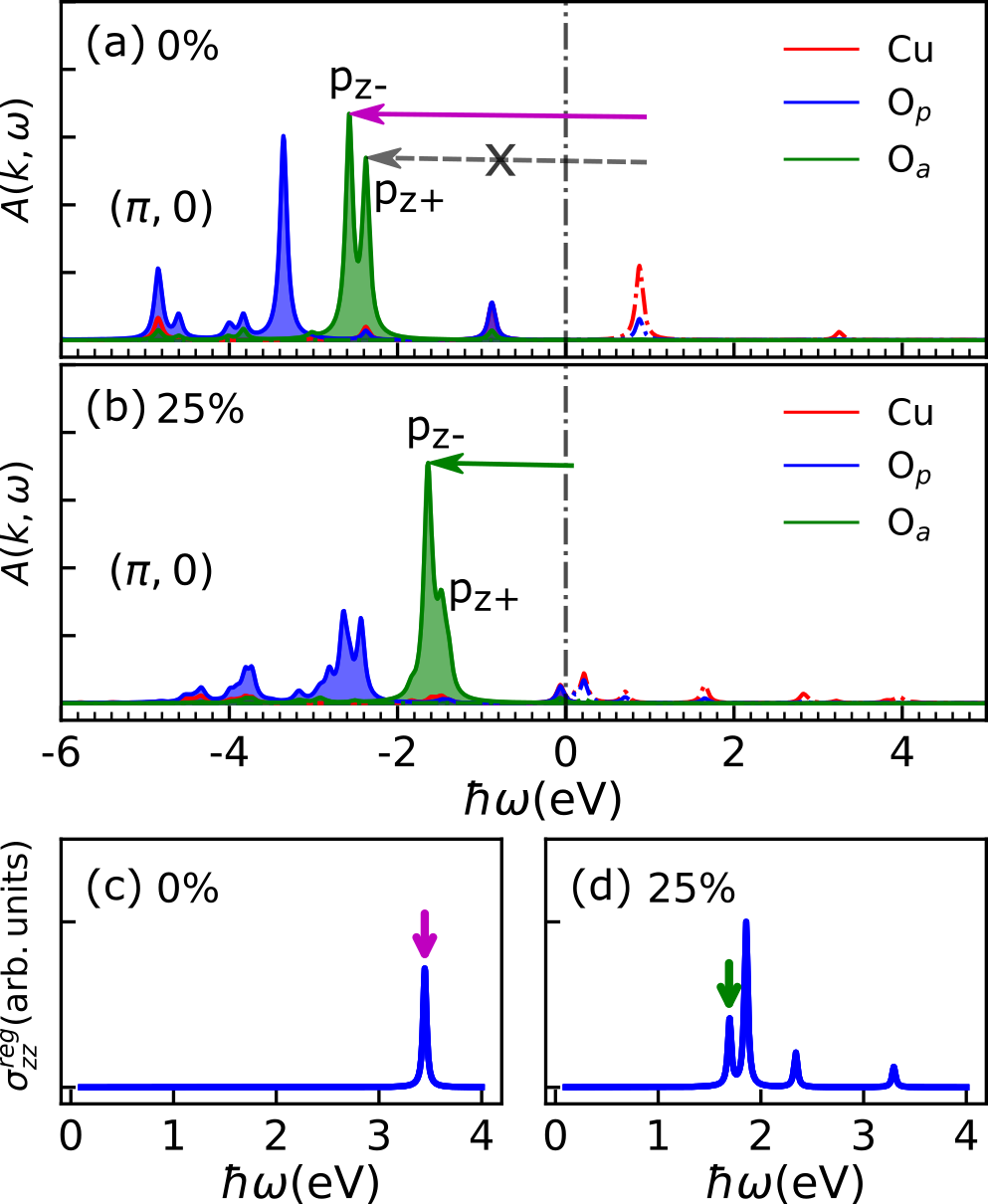}
  \end{center}
  \caption{
    (a) and (b) are orbital-resolved single-particle spectral functions at momentum ($\pi$, 0) for 0\% and 25\% dopings respectively. 
    The red line represents copper orbitals (denoted by Cu), the blue line denotes in-plane oxygen orbitals (O$_p$), and the green line represents apical oxygen orbitals (O$_a$). 
    The shaded (dashed) curves represent hole additional (removal) spectral functions. The vertical dashed line denotes the middle of the highest occupied state and the lowest unoccupied state. 
    The arrows in (a) and (b) mark possible transitions between in-plane bands and apical oxygen bands.
    (c) and (d) are regular part of of the $c$-axis optical conductivity for 0\% and 25\% dopings, respectively. 
  }
  \label{pic:spec}
\end{figure}

Possible $c$-axis excitations are inferred from the equilibrium single-particle spectral function $A(k,\omega)$ and optical conductivity $\sigma_{zz}^{reg}(\omega)$.
Here, we only show $A(k,\omega)$ at the momentum point ($\pi$, 0) (see FIG.~\ref{pic:spec}(a)(b)) which lies close to the Fermi level and provides the lowest energy excitations between the in-plane orbitals and apical oxygen
\footnote{To ensure that our model evaluated on such a small cluster produces a reasonable distribution of spectral weight, we have examined our single-particle spectral functions and find they are consistent with previous numerical and experimental results. We find consistency between the spectral weight for Cu and in-plane oxygen orbitals and that from previous studies of an in-plane model\cite{CHEN2011106}; at half-filling the in-plane charge transfer (CT) gap between the upper Hubbard band (UHB) and the Zhang-Rice singlet (ZRS) band is $\sim1.8$ eV, matching optical measurements\cite{PhysRevB.41.11657, PhysRevB.43.7942}; and the position of the apical oxygen spectral weight qualitatively agrees with that from \emph{ab initio} calculations, and also matches rather well to information extracted from ARPES measurements\cite{PhysRevB.99.224509}}.

There are two prominent apical oxygen states labeled as $p_{z-}$ and $p_{z+}$ having parity $\mp$, respectively. At half-filling shown in FIG.~\ref{pic:spec}(a), possible hole transitions from the UHB to $p_{z-}$ and $p_{z+}$ are labeled by magenta and grey arrows. However, only transitions to $p_{z-}$ at 3.4 eV appear in $\sigma_{zz}^{reg}(\omega)$, as shown in FIG.~\ref{pic:spec}(c). 
In the doped case, holes near the Fermi level are predominantly Zhang Rice Singlet (ZRS) states, and the allowed excitations at low energies involve transitions to the $p_{z-}$ band [labeled by the green arrow in FIG.~\ref{pic:spec}(b)], which also has a correspondence in  $\sigma_{zz}^{reg}(\omega)$, as shown in FIG.~\ref{pic:spec}(d). The excitation energy is reduced from 3.4 eV to 1.7 eV compared to half-filling due to the different nature of the transition. 

These results directly show that resonant photoexcitation at these energies are clearly connected to creating holes in the symmetry allowed apical oxygen states. 

\begin{figure}[b!]
  \begin{center}
    \includegraphics[width=8.5cm]{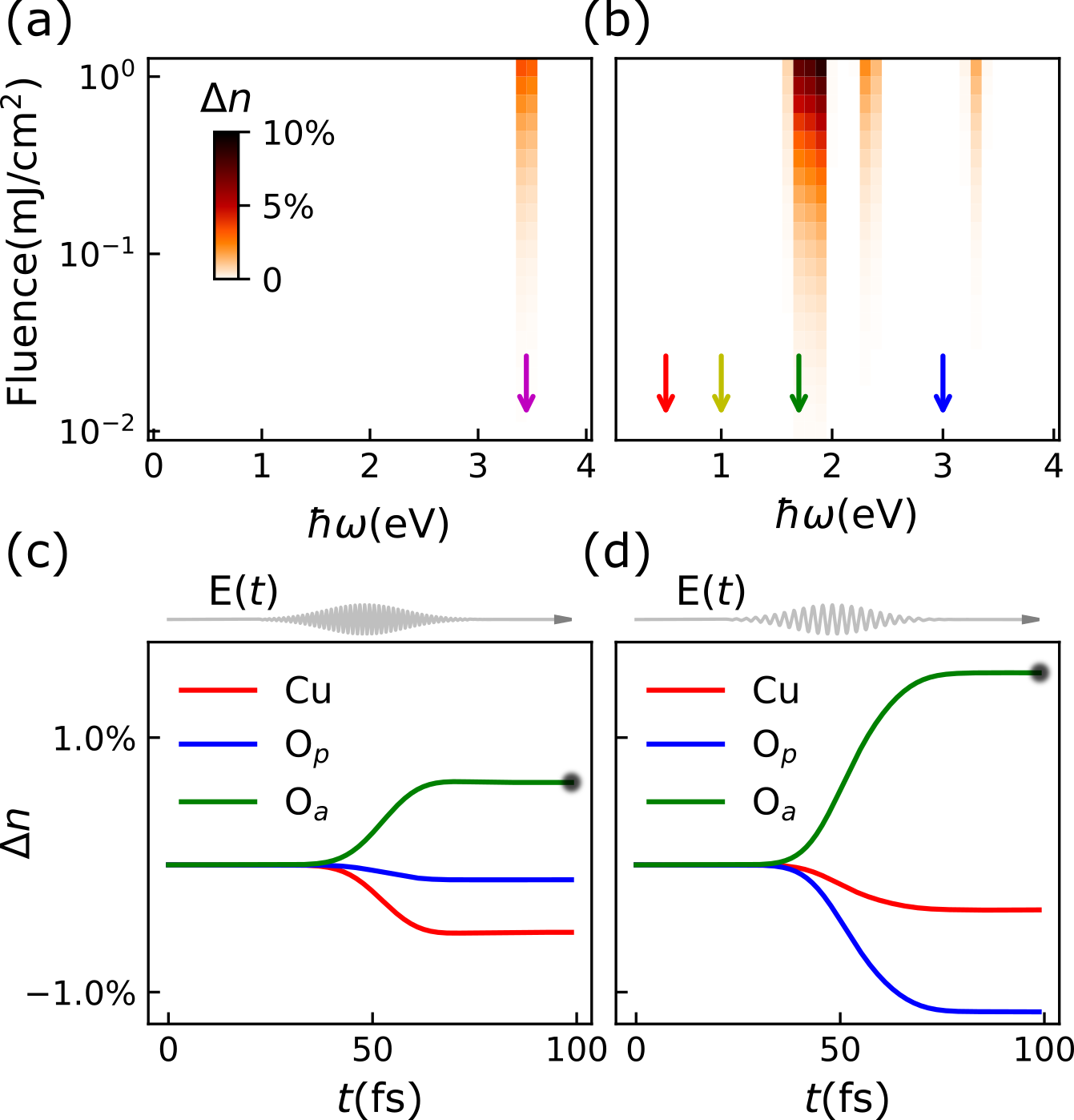}
  \end{center}
  \caption{
    The color map indicates density change $\Delta n(t_f)$ on apical oxygen due to holes transferred to apical oxygen $p_z$ orbitals form in-plane orbitals after being excited by pulses with different frequencies and fluences. (a) and (b) are results for 0\% and 25\% dopings respectively.
    The magenta arrow in (a) and green arrow in (b) annotate the lowest allowed transitions in each doping case, 
    and correspond to the arrows with the same color in FIG.~\ref{pic:spec}(a) and (b).
    (c) Time evolution of density change on different orbitals. 
    The $c$-axis polarized pump pulse $E(t)$ with frequency 3.4eV and fluence $0.2mJ/cm^2$ is shown as grey line. 
    This frequency is marked by the magenta arrow in (a).  
    (d) Similar to (c) but for the 25\% doped case and the pump frequency is 1.7eV, which is annotated by the green arrow in (b). The black dots in (c) and (d) mark $\Delta n(t_f)$ on apical O$_a$ orbitals.
  }
  \label{pic:scan}
\end{figure}

We now study the non-equilibrium dynamics induced by a $c$-axis polarized pulse, which is represented by an effective electric field polarized along the $c$-axis with a pulse shape given by
\begin{equation}\label{eq:pulse}
 \mathbf{E}(t) = E_0\exp\left(-\frac{(t-t_0)^2}{2\sigma^2}\right)\sin(\omega(t-t_0) + \phi)\mathbf{e_z},
\end{equation}
where $E_0$ is the amplitude, $t_0$ is the center of the pulse, $\sigma$ is the pulse width, $\omega$ is the angular frequency, $\phi$ is the pulse phase and $\mathbf{e_z}$ is a unit vector along the $c$-axis.
To mimic experiments, where pulse trains will have varying phases, we also average over the pulse phase $\phi$ when performing time evolutions. The vector potential used in the Peierls substitution is obtained by integrating $E(t)$ over time. And the fluence of the pump is determined by  $F = \int_{-\infty}^{\infty} c\epsilon_0 |E(t)|^2 dt$. 

To fully characterize the pump frequency and fluence dependence of the $c$-axis excitations, we apply $c$-axis polarized pump fields with different frequencies and intensities to the ground state obtained for both undoped (0\% doped-holes) and 25\% doping, respectively. The pump pulse is centered on a window in time from $t_i = 0$\,fs to $t_f=100$\,fs. The field strength is negligibly small at $t_i$ and $t_f$. We measure the change in hole density on different orbitals, defined as 
$\Delta n_\mu(t) = [N_\mu(t) - N_\mu(t_i)]/4\times 100\%$, 
where $\mu$ denotes the orbital index and $N_\mu(t)$ is the number of holes in orbital $\mu$ at time $t$. For convenience, we will use Cu for $3d_{x^2-y^2}$ orbitals, O$_p$ for the sum of $2p_x$ and $2p_y$ orbitals, and O$_a$ for the sum of the two $2p_{z}$ orbitals.

We can gauge the effectiveness of photodoping by measuring how many holes are transferred to apical oxygens at time $t_f$, which is proportional to $\Delta n_{\textrm{O}_a}(t_f)$ (see FIG.~\ref{pic:scan}(a)(b)).
At half-filling, the external pump cannot efficiently induce photodoping until the frequency is resonantly tuned to the resonant value 3.4eV (magenta arrow in FIG.~\ref{pic:scan}(a)).
While there are multiple resonances in the doped system, for our cluster the lowest resonance occurs at 1.7eV (green arrow in FIG.~\ref{pic:scan}(b)) corresponding to the transition from ZRS to $p_{z-}$ states. This energy is close to the frequency where experiments show the largest changes in the optical conductivity, consistent with promoting holes from the copper-oxygen plane into apical oxygen states. 

However, photodoping does not necessarily give similar effects to chemical doping, and as such, enables the shift of the superconducting phase. A necessary condition would be that the distribution of photoinduced carriers among different (in-plane) orbitals be similar to that of chemical doping. We note that such correspondence generally would not hold, since photoexcitation usually causes inplane charge distributions to deviate from the equilibrium distributions. 
To check if this correspondence holds, we examine the evolution of hole density at the lowest resonance for each doping. As shown in FIG.~\ref{pic:scan}(c), we find that the 3.4 eV pulse, in a half-filled system, primarily transfers holes from Cu to O$_a$, which results in a charge distribution that deviates from equilibrium chemical doping\cite{PhysRevB.93.155166}. 
In contrast, when we excite the 25\% doped system by resonant pump (i.e.~the 1.7 eV pulse), holes are transferred primarily from in-plane O$_p$ to O$_a$, leading to a hole distribution consistent with chemical doping [see Fig.~\ref{pic:scan}(d)]. More quantitatively, the ratio between $\Delta n_{\textrm{O}_p}(t_f)$ and $\Delta n_{\textrm{Cu}}(t_f)$ is about 3.26:1, closely matching the ratio (3.25:1) for the introduction of carriers in equilibrium\cite{PhysRevB.93.155166}. Such an agreement may mean that $c$-axis photodoping would lead to similar effects as chemical doping, in which doped-holes reside primarily on in-plane oxygen orbitals, and the effects can be mapped to the phase diagram in Fig.~\ref{pic:lattice}(a) in a doped system.

Now, we focus on the 1.7\,eV pump in the 25\% doped system. After the pump with a fluence of $0.2$\,mJ/cm$^2$ comparable to those used experimentally\cite{PhysRevB.91.174502}, the in-place doping concentration is reduces by $\sim 1.5\%$.
If we continue to increase the field strength with the same resonant frequency, more holes would be transferred to apical oxygens. In the meantime, the initial states will get depleted and the final states will get filled, thus we expect to observe a saturation, which also is present in experiment\cite{PhysRevB.91.174502}.

\begin{figure}[t!]
  \begin{center}
    \includegraphics[width=8.5cm]{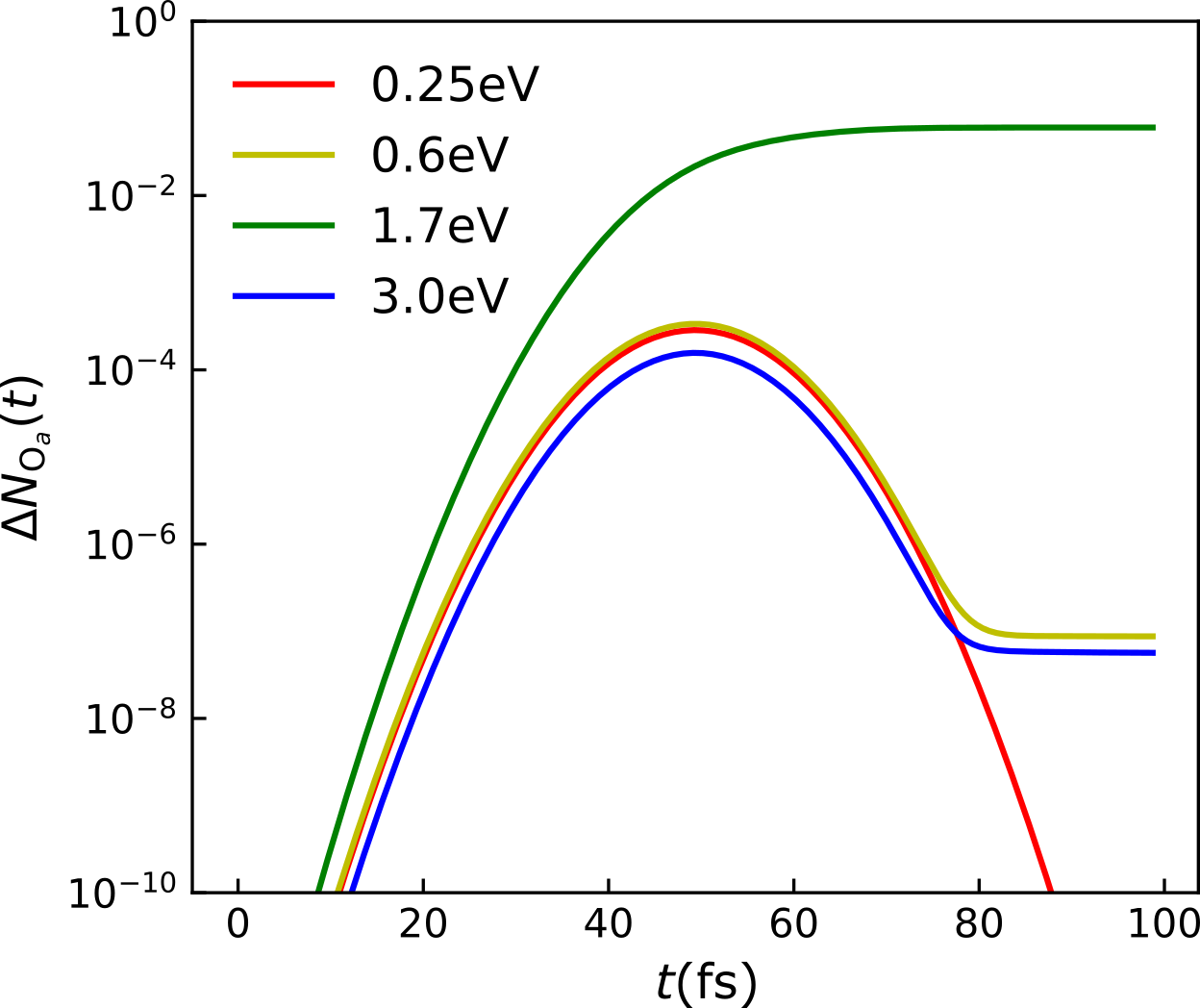}
  \end{center}
  \caption{ 
    We plot $\Delta N_{\textrm{O}_a}(t) = N_{\textrm{O}_a}(t) - N_{\textrm{O}_a}(t_i)$ when excited by $c$-axis polarized pulse. $N_{\textrm{O}_a}(t)$ is the number of holes on apical oxygen orbitals at time $t$. 
    The different pulse frequencies correspond to those indicated by the arrows in FIG.~\ref{pic:scan}(b) and the pulse fluence is set to be $0.2mJ/cm^2$. 
    In order to see the small changes induced by the 0.25eV, 0.6eV and 3.0eV pump pulses, we used a log scale for $\Delta N_{\textrm{O}_a}(t)$.
  }
  \label{pic:resonance}
\end{figure}

To highlight further the wavelength selectivity for $c$-axis excitation, we choose three additional photon energies 0.25 eV, 0.6 eV and 3.0 eV as roughly the same used in experiments\cite{PhysRevB.91.174502}.  We compare $\Delta N_{\mathrm{O}_a}(t)$ for photoexcitation at these frequencies to that for the 1.7eV pulse in FIG.~\ref{pic:resonance}. We note that the sharp $c$-axis excitations obtained in the ED study presented here will broaden with a width depending on the band width of the ZRS and $p_{z-}$ bands in the thermodynamic limit. We would expect to observe such a photodoping effect in a window of energy around the resonances associated with photoexcitation between the ZRS and $p_{z-}$ states.

\section{Discussion and summary}
We note that this small cluster used in ED certainly can not elucidate stripe or superconducting phases\cite{PhysRevLett.80.1272, PhysRevB.60.R753, PhysRevLett.113.046402, Zheng1155, Jiang1424, Huang1161}. But direct numerical signatures for equilibrium superconductivity alone, not even addressing photo-induced superconductivity out-of-equilibrium, remain an extremely hard problem, and even then the information comes in the form of a direct pairing correlation function and not a signature in a simulated optical response. Finding signature of superconductivity is certainly beyond the capability of exact diagonalization for a study which involves multiple orbitals and out-of-equilibrium time evolution.
However ED remains a valuable method to provide preliminary insights and has some advantages over other numeric methods: it allows easy calculation of dynamic quantities such as single-particle spectral function, which matches experiments rather well; it is one of the most powerful numeric methods to study long period time evolution, which enables us to study c-axis excitation. Furthermore, as this study focuses on photon selectivity along the $c$-axis, the charge transfer to apical orbitals is much more pronounced than that of the stripe/superconducting order. We believe ED is appropriate to study c-axis charge transfer and to help contemplate how this can affect superconductivity.

Our results indicate that a purely electronic-based resonant photoexcitation of holes from planar to apical states can be actuated via $c-$axis light polarization in an orbitally dependent way. For reasonable pulse strengths used in experiments\cite{PhysRevB.91.174502}, we see that photodoped holes on the order of a percent may be transferred into apical states, shifting and reducing the number of planar hole carriers away from the 1/8 stripe anomaly where superconductivity is more robust in equilibrium. Therefore we conjecture that photodoping is sufficient to describe the changes in optical conductivity seen in Ref. \cite{PhysRevB.91.174502}. 

This picture of $c$-axis polaized pump-induced photodoping likely applies beyond the 1/8 anomaly, and can be further exploited in different material classes. 
Just from the point of photodoping to remove holes in the CuO plane, we would predict that $c$-axis photodoping should enhance superconductivity in the over-doped region, but suppress it for under-doping. However, compared to near the 1/8 anomaly where the outsized effect of charge-stripe order gives a dramatic doping dependence of $T_c$ , we may expect that the effect of reduced in-plane doping alone may only lead to small changes in the superconducting transition temperature in other doping regions. The overall effect of light induced charge redistribution in other doping regions may thus require a more careful analysis.

We also expect similar photodoping effects involving transitions from the ZRS band to apical $p_z$ band for cuprates with one apical oxygen per copper atom\cite{PhysRevX.10.011053}, but without the selection rule imposed by the mirror symmetry.
However, for materials without apical oxygens such as CaCuO$_2$, the apical Ca orbitals weakly couple to the in-plane orbitals, and have negligibly small density of states in a large energy window ($>8$ eV) below the Fermi level\cite{PhysRevX.11.011050}. The $c$-axis excitation from in-plane to apical orbitals, if possible, will thus require pulses with much higher photon energies far outside the optical range.

We stress that photodoping typically cannot be mapped simply to an equilibrium chemical doping. For example, the conclusions of this paper cannot be extended to in-plane polarized pulses, since in such cases holes move between in-plane O and Cu orbitals, in addition to apical charge transfer. Therefore, the nonequilibrium in-plane hole distribution will deviate significantly from any equilibrium doping and \emph{cannot} be addressed in the guise of a photodoping. A sizeable simulation of in-plane competing orders and their response to the ultrafast pump would be required in such cases.

The idea of using light to manipulate charge distribution among selective orbitals and thus influence underlying orders can be extended beyond cuprates. Natural extensions include perovskite structures ABX$_3$ or other transition metal compounds, where the metal atoms are coordinated by in-plane and apical ligands.
For specific examples, one could use c-axis polarized light to transiently change the in-plane doping levels of manganites and control the underlying spin, charge and orbital orders\cite{RevModPhys.73.583, Tokura462};
or in SrTiO$_{3-\delta}$, light could be used to transiently alter the superconducting state, shifting the band filling near the critical doping level, where the increase of $T_c$ in the dilute limit, initial filling of the lowest Ti 3d band, is interrupted as soon as the middle 3d band appears\cite{doi:10.1146/annurev-conmatphys-031218-013144}.

Finally, we point out that time-resolved X-ray absorption spectroscopy\cite{Bressler:2004aa, PhysRevB.101.165126} can be used to monitor hole distribution among different orbitals when driven by $c$-axis polarized fields as an experimental validation of this photodoping effect. 

\section*{Acknowledgement}
The authors would like to thank Yu He for helpful discussion, and Daniele Nicoletti for helpful comments and
suggestions. This work was supported by the U.S. Department of Energy, Office of Basic Energy Sciences, Division of Materials Sciences and Engineering, under contract DE-AC02-76SF00515. The computational results utilized the resources of the National Energy Research Scientific Computing Center (NERSC) supported by the U.S. Department of Energy, Office of Science, under Contract No. DE-AC02-05CH11231.

\bibliography{main}
\end{document}